\documentclass{article}  
\usepackage{bigsky2009}
\usepackage{graphicx}
\frompage{000} \topage{000}                                              
\def\rightmark{Direct Photons in PHENIX}
\def\leftmark{S. Bathe et al.}
 
\title{Enhanced Direct Photon Production in Au+Au Collisions at 
$\sqrt{s_{\scriptscriptstyle NN}}$ = 200 GeV in PHENIX} 
\authors{
{S. Bathe$^1$ for the PHENIX collaboration %
}\\[2.812mm]
{\normalsize
\hspace*{-8pt}$^1$ RIKEN-BNL Research Center \\ 
Brookhaven National Laboratory\\
Upton NY 11973, USA\\[0.2ex] 
}}
 
\abstract{The production of electron pairs with $p_T$
between 1 and 5 GeV/c and $m<$ 300 MeV has been measured at
mid-rapidity in $\sqrt{s_{\scriptscriptstyle NN}}$ = 200 GeV $p+p$ and Au+Au collisions by
the PHENIX experiment at RHIC.  A significant excess above the
hadronic background was observed in both $p+p$ and Au+Au collisions.
Treating the excess as internal conversion of direct photons, the
direct photon yield in Au+Au was found to be enhanced compared to the
binary-scaled $p+p$ yield.  The enhancement is consistent with an
exponential inverse slope of $221 \pm 23 \pm 18$ MeV and predictions
from hydrodynamical models with initial temperature between 300 and
600 MeV at formation times of 0.6--0.15 fm/c.  }

 
\begin{document}
 
\maketitle
\setcounter{page}{1}

\section{Introduction}\label{intro}
\begin{figure}[htb]
\begin{minipage}[t]{0.475\linewidth}
\vspace*{-.2cm}
                 \includegraphics[width=\linewidth]{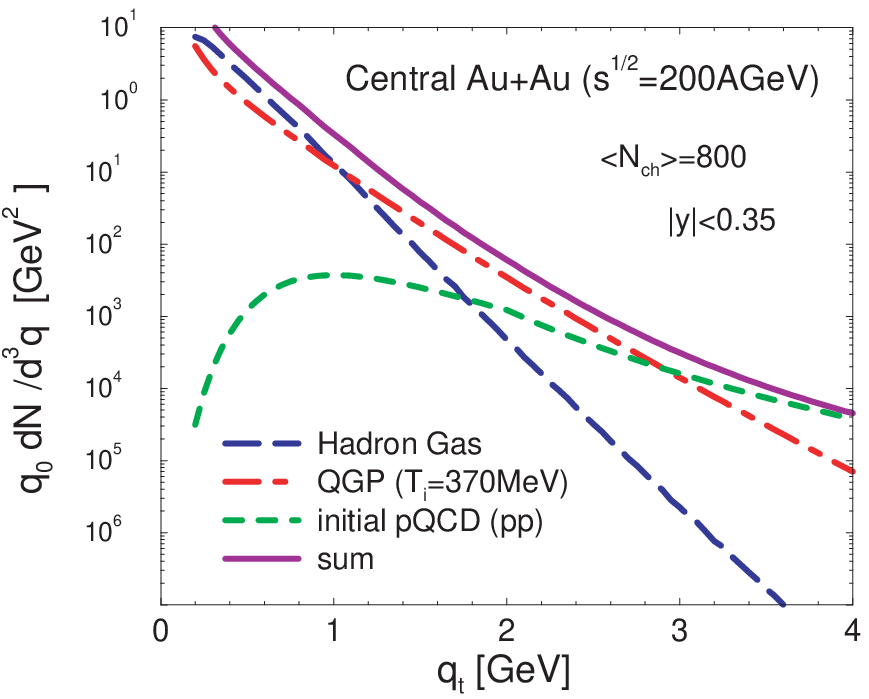}
\vspace*{-1cm}
\caption[]{Direct photon sources at RHIC \cite{Turbide:2003si}.}
\label{fig:turbide}
\end{minipage}
\hspace{\fill}
\begin{minipage}[t]{0.475\linewidth}
\vspace*{-.2cm}
                 \includegraphics[width=\linewidth]{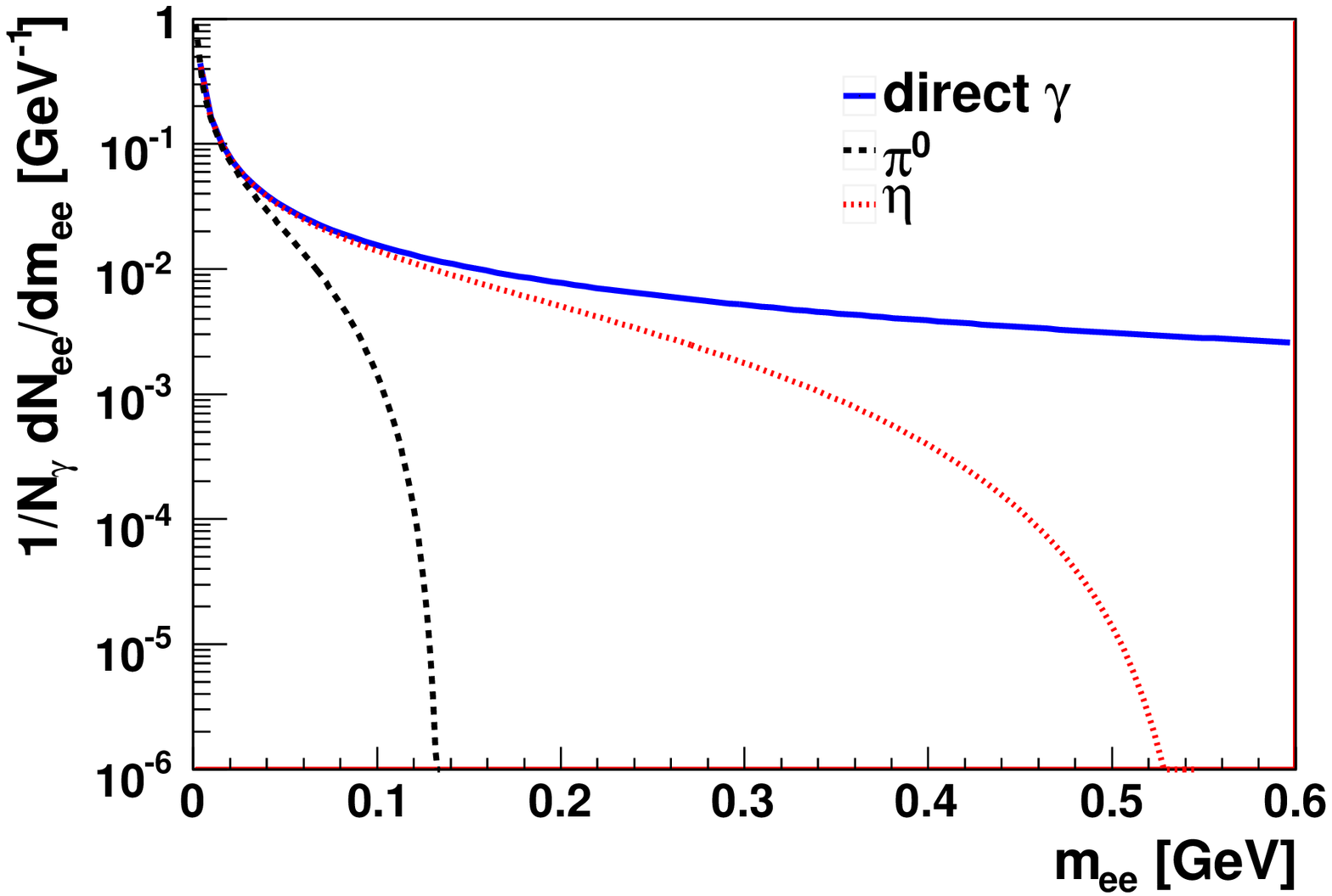}
\vspace*{-1cm}
\caption[]{Illustration of the pair yield mass dependence for direct
photons as well as for $\pi^0$ and $\eta$ Dalitz pairs.}
\label{fig:gpe}
\end{minipage}
\end{figure}

A multitude of results from the Relativistic Heavy Ion Collider (RHIC)
at Brookhaven National Laboratory (BNL) indicates the creation of a
high-density, thermalized medium in ultra-relativistic collisions of
heavy ions \cite{Adcox:2004mh}.  Such a medium is expected to radiate
thermal photons \cite{Turbide:2003si}, which, once produced, leave the medium
unscathed.  Thermal photons from the partonic phase of the collision
are predicted to dominate the direct photon spectrum in the transverse
momentum ($p_T$) range of 1--3 GeV/c as illustrated in Fig. \ref{fig:turbide} \cite{Turbide:2003si}.
However, here direct photons are submerged in a
background of hadronic decay photons, mainly from the $\pi^{0}$ and
$\eta$.  This background constitutes a major experimental challenge in
the conventional, calorimeter-based measurement.  It can be overcome,
though, by measuring low-mass electron pairs in a mass range
where electron pairs from the $\pi^{0}$ Dalitz decay do not
contribute \cite{Adare:2008fqa}.  The electron-pair yield above the hadronic background can
be treated as internal conversion of direct photons \cite{Cobb:1978gj}.

\section{Internal Conversion Method}\label{ICA}  
Any source of high-energy photons also emits virtual photons with very
low mass \cite{Cobb:1978gj}.  Those virtual photons then convert to low-mass
$e^{+}$--$e^{-}$ pairs, which can be measured ({\it Internal Conversion
Method}).  The pair yield per direct photon falls with the pair mass
as:
\begin{eqnarray}
\frac{d^2n_{ee}}{dm} = \frac{2\alpha}{3\pi}\frac{1}{m} \sqrt{1-\frac{4m_e^2}{m^2}}\Bigl( 1+\frac{2m_e^2}{m^2} \Bigr) S  dn_{\gamma} \label{eq:Conversion}.
\end{eqnarray}
Here $\alpha$ is the fine structure constant, $m_e$ and $m$ are the
masses of the electron and the $e^+$--$e^-$ pair, respectively, and $S$ is
a process dependent factor that goes to 1 as $m \rightarrow 0$ or $m
\ll p_T$.  Equation \ref{eq:Conversion} also describes the relation
between photons from hadron decays (e.g. $\pi^0, \eta \rightarrow
\gamma \gamma$, and $\omega \rightarrow \gamma \pi^0$) and $e^+$--$e^-$
pairs from Dalitz decays ($\pi^0, \eta \rightarrow e^+$--$e^-\gamma$ and
$\omega \rightarrow e^+$--$e^-\pi^0$).  For $\pi^0$ and $\eta$, the factor
$S$ is given by $S =|F(m^2)|^2
(1-\frac{m^2}{M_h^2})^3$~\cite{Landsberg:1986fd}, where $M_h$ is the
meson mass and $F(m^2)$ is the form factor.  Figure \ref{fig:gpe}
illustrates the pair yield mass dependence for direct photons as well
as for $\pi^0$ and $\eta$ Dalitz pairs.  The cut-off of the $\pi^0$
pairs at the $\pi^0$ mass can be exploited to increase the
signal-to-background ratio from 10\%, where it is comparable to the
systematic uncertainty and therefore not significant, to 50\%, making
a significant measurement possible.  Since the measurement at low
$p_T$ is systematics limited, the simultaneous reduction in
statistical significance is an acceptable trade-off\footnote{There are
about 0.001 virtual photons with $m_{ee} > M_{\pi^{0}}$ for every real
photon.}.

\section{Data Set And Backgrounds}\label{data}  
The analysis is based on two data sets: Au+Au at $\sqrt{s_{\scriptscriptstyle NN}}$ = 200
GeV acquired in 2004 consisting of 0.8 billion minimum bias events
(4.9 pb$^{-1}$ $p+p$ equivalent); $p+p$ at the same cms energy acquired in
2005 with 2.25 pb$^{-1}$.  Charged tracks were measured with the Drift
Chamber and Pad Chamber of the PHENIX \cite{Adcox:2003zm} Central Arms covering
$|\eta|<0.35$ and $\Delta \phi = 2 \times \pi/2$ and identified as
electrons with the Ring Imaging \v{C}erenkov Detector (RICH) and
Electromagnetic Calorimeter (EMCal).  Material conversion pairs were
removed by a cut on the orientation of the pair plane with respect to
the magnetic field.  Combinatorial background was removed by an event
mixing technique, accurate to 0.25\% systematic uncertainty in Au+Au.

There is additional correlated background from two sources: {\it cross
pairs} with one electron/positron from either virtual photon in a
double-Dalitz decay; {\it jet pairs} from two different Dalitz decays
within the same jet or from back-to-back jets.  These contributions
can be well understood in a Monte Carlo calculation and have been
subtracted.

\section{Signal Extraction}\label{signal}  
After subtraction of the combinatorial background and the cross and
jet pairs, the pair mass spectrum is compared to a ``cocktail'' of
\begin{figure}[htb]
\vspace*{-.2cm}
                 \insertplot{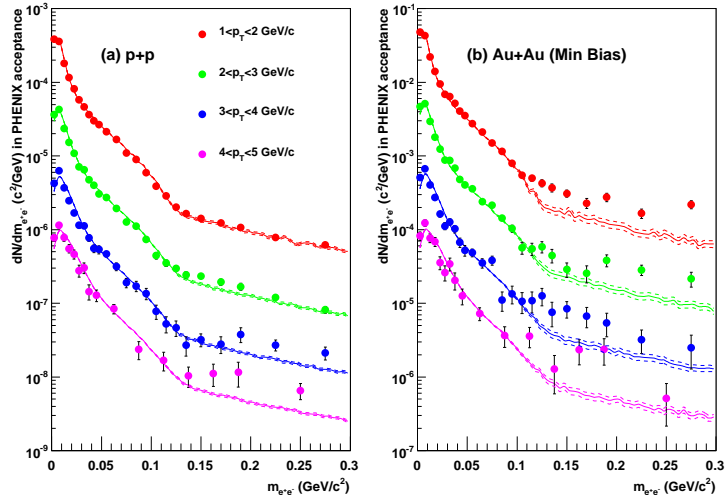}
\vspace*{-1cm}
\caption[]{Pair mass spectra for data (points) and hadronic cocktail
(lines) in the PHENIX acceptance for different $p_T$ intervals for $p+p$
(left) and Au+Au (right)~\cite{Adare:2008fqa}.}
\label{fig:mass}
\end{figure}
known hadronic sources \cite{Afanasiev:2007xw,Adare:2008asa}.  Figure \ref{fig:mass}
shows this comparison for different $p_T$ intervals for both $p+p$ and
Au+Au.  The cocktail is normalized to the data for $m<30$ MeV, where
the $\pi^{0}$ Dalitz decay dominates the yield.  The ``knee'' at 100
MeV comes from the $\pi^{0}$ cut-off leading to the 80\% background
reduction mentioned above.

In $p+p$, the pair yield is consistent with the hadronic background for
the lowest $p_T$ interval.  At higher $p_T$ a small excess is visible
for $m>m_{\pi^{0}}$.  In Au+Au a much larger excess appears at all
$p_T$, indicating an enhanced production of virtual photons\footnote{This excess is in a different kinematic region (higher $p_T$,
lower mass) than the low-mass enhancement reported in
\cite{Afanasiev:2007xw}, which is expected to be dominated by the
hadronic gas phase.}.

\begin{figure}[htb]
\vspace*{-.2cm}
                 \insertplot{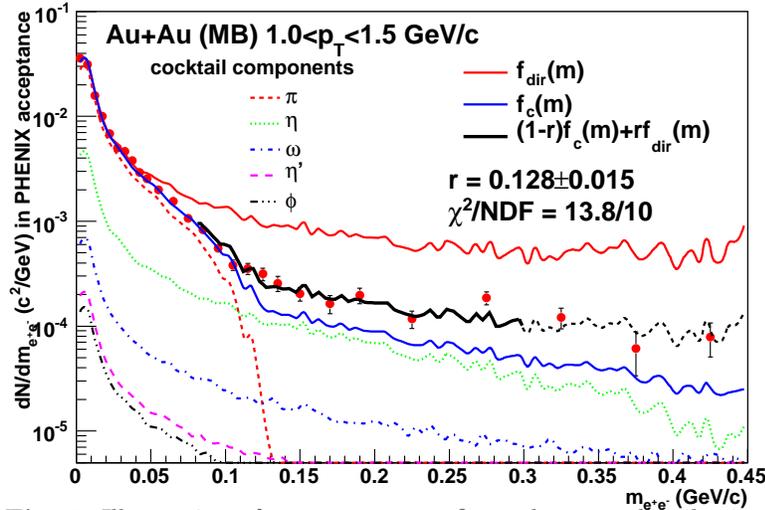}
\vspace*{-1cm}
\caption[]{Illustration of two-component fit to the mass
distribution \cite{Adare:2008fqa}.}
\label{fig:fit}
\end{figure}

To quantify the direct photon fraction, the mass spectrum is fit with
a two-component function $f(m_{ee}) = (1-r)f_{c}(m_{ee}) + r
f_{dir}(m_{ee})$ as illustrated in Fig.~\ref{fig:fit}. Here
$f_{c}(m_{ee})$ is the shape of the cocktail mass distribution (shown
in Fig.~\ref{fig:mass}), $f_{dir}(m_{ee})$ is the expected shape of
the direct photon internal conversion, and $r$ is the fit parameter.
Both $f_{c}(m_{ee})$ and $f_{dir}(m_{ee})$ are separately normalized
to the data for $m_{ee}<30$ MeV/$c^2$, where their shapes are nearly
identical.  This preserves the meaning of $r$ as the {\it real} direct
photon fraction.
From the agreement between data and fit it can be concluded that the
data matches the expected shape for direct photons.

\begin{figure}[htb]
\vspace*{-.2cm}
                 \insertplot{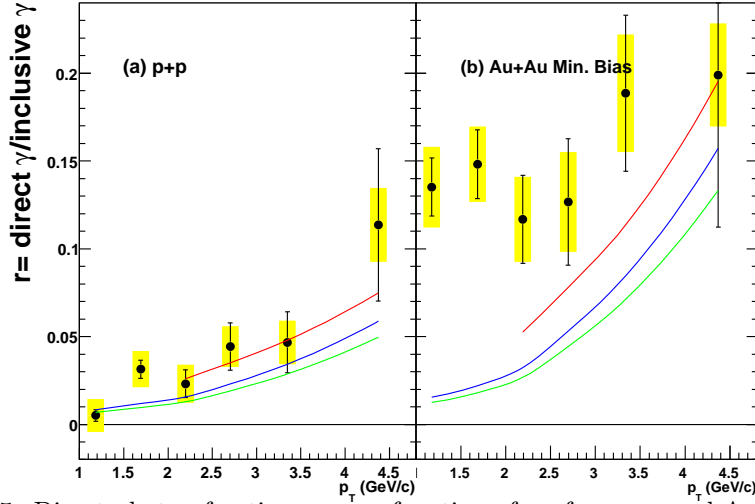}
\vspace*{-1cm}
\caption[]{Direct photon fraction, $r$, as a function of $p_T$ for $p+p$
and Au+Au events compared to a pQCD calculation for three different
scales.  In the case of Au+Au the pQCD result is scaled by the nuclear
overlap function, $T_{AA}$ \cite{Adare:2008fqa}.}
\label{fig:frac}
\end{figure}

The so extracted direct photon fraction, $r$, is plotted as a function
of $p_T$ in Fig.~\ref{fig:frac}.
While in case of $p+p$ the direct photon fraction is consistent with
pQCD, for Au+Au $r$ is enhanced above pQCD.

As this measurement is based on shape differences to extract the
direct photon fraction, the $\eta/\pi^{0}$ ratio is the largest source
of systematic uncertainty.  This results in a 7\% uncertainty in $p+p$
and 17\% in Au+Au.  Other sources contribute only a few percent as
the cocktail is normalized to the data and no absolute normalization
is required.

In the next step, $r$ is converted into the direct photon yield as
$dN^{dir}(p_{\rm T})= r \times dN^{incl}(p_{\rm T})$.  The inclusive
photon yield for each $p_{\rm T}$ bin is determined by
$dN_\gamma^{incl}=N_{ee}^{data} \times (dN_{\gamma}^{c}/N_{ee}^{c})$,
where $N_{ee}^{data}$ and $N_{ee}^{c}$ are the measured and the
absolutely normalized cocktail $e^+$--$e^-$ pair yields, respectively,
both for $m_{ee}<30$ MeV/$c^2$; and $dN_{\gamma}^{c}$ is the yield of
photons from the cocktail.  Here we use the fact that the ratio of the
photon yield to the $e^+$--$e^-$ pair yield for $m_{ee}<30$ MeV/$c^2$
calculated from Eq.~\ref{eq:Conversion} is the same within a few
percent for any photon source.  The systematic uncertainty of
$\gamma^{incl}$ is 14\% from the $e^+$--$e^-$ pair acceptance.

\section{Results}\label{results}  
Figure \ref{fig:yield}
shows the invariant yield of direct photons as a function of $p_T$ for
$p+p$ and three different centrality classes in Au+Au.  The $p+p$ data is
again compared to the pQCD calculation.  The calculation is consistent
with the data for $p_T>2$ GeV.  A modified power law, $A_{pp}
(1+p_{T}^2/b)^{-n}$, fits the data over the entire $p_T$ range.

\subsection{Significance of the Modified Power Law}\label{powerlaw}  
\begin{figure}[htb]
\begin{minipage}[t]{0.475\linewidth}
\vspace*{-.2cm}
                 \includegraphics[width=\linewidth]{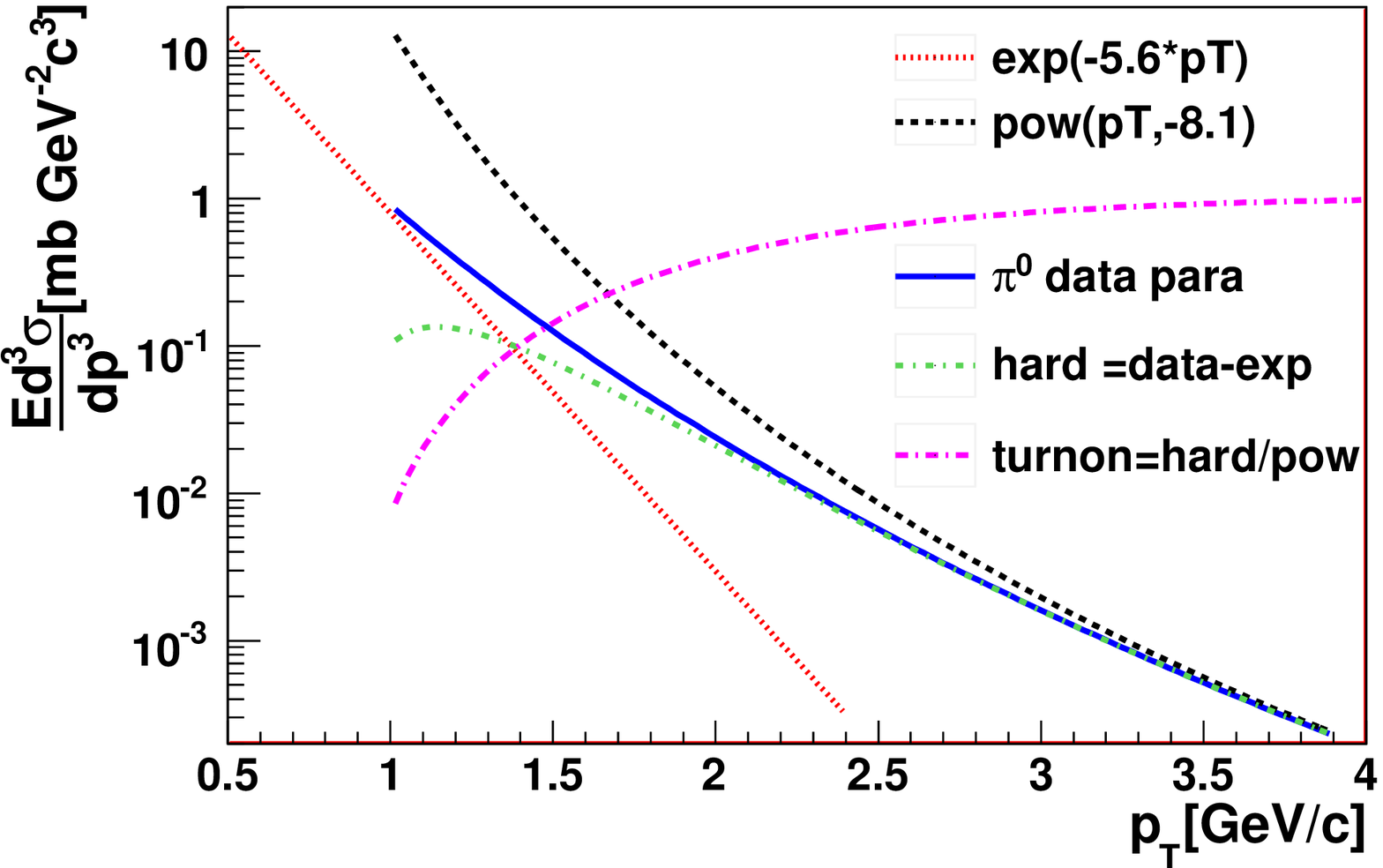}
\vspace*{-1cm}
\caption[]{Parameterization of $\pi^{0}$ production in $p+p$
\cite{Adler:2006wg} and its various contributions (for details see text).}
\label{fig:hardBreakDown}
\end{minipage}
\hspace{\fill}
\begin{minipage}[t]{0.475\linewidth}
\vspace*{-.2cm}
                 \includegraphics[width=\linewidth]{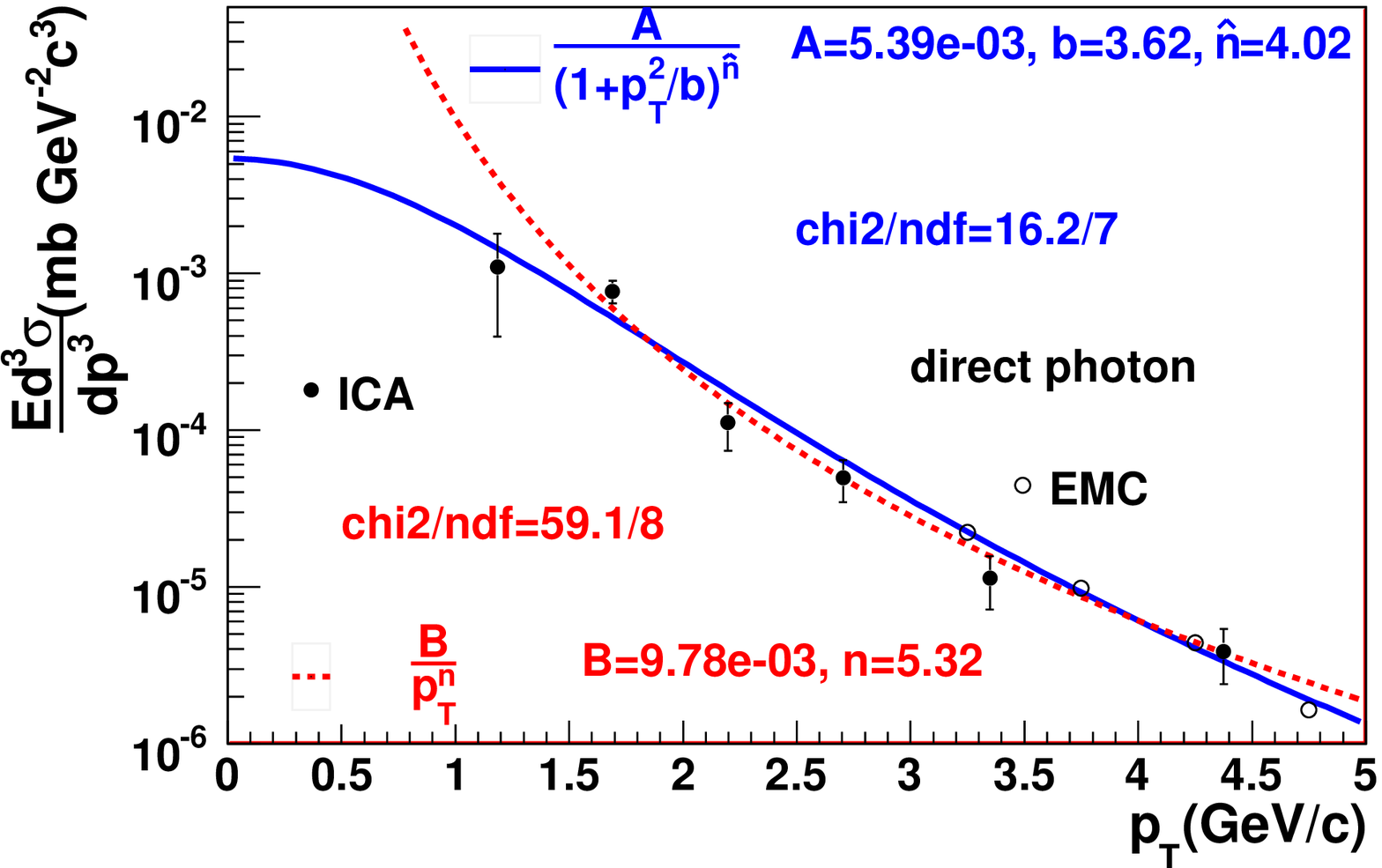}
\vspace*{-1cm}
\caption[]{Fit of direct photon cross section with both a power law and a modified power law.}
\label{fig:twoFits}
\end{minipage}
\end{figure}

The modified power law appears to yield an at least as good
description of the data at low $p_T$ as the pQCD calculation.  What is the significance of this observation?  It is
obvious that the power-law behavior of hard scattering has to break
down for $p_T \rightarrow 0$.  For hadrons, soft production sets in
with an exponential slope.  This is illustrated in
Fig. \ref{fig:hardBreakDown},
which shows a parameterization of $\pi^{0}$ production in $p+p$~\cite{Adler:2006wg}.
The parameterization is the sum of a power law and an exponential with
a Woods-Saxon transition between the two.  The two functions are shown separately.  The exponential dies out at
high $p_T$ while the power law diverges at low $p_T$.  The
hard-scattering contribution can be understood as the difference
between the data parameterization and the exponential contribution.  One can see that
it flattens out towards low $p_T$, deviating from the power law.  This
flattening corresponds to an onset of hard scattering as illustrated
by the ratio of the hard scattering contribution to the power law.

This onset is not directly observable for hadrons as the production at
low $p_T$ is dominated by soft physics.  Direct photons, however, are
only produced in hard scatterings.  This makes the onset of hard
scattering at low $p_T$ directly measurable.  The onset has also been measured
in Drell-Yan production \cite{Ito:1980ev}.

To evaluate the statistical significance of the onset, the PHENIX data
were fitted with both a power law and a modified power law.  As shown
in Fig. \ref{fig:twoFits}, the modified power law yields a smaller
reduced $\chi^2$ than the pure power law.

\subsection{Au+Au Enhancement}\label{enahnced}  
\begin{figure}[htb]
\vspace*{-.2cm}
\begin{center}

                 \includegraphics[width=0.7\linewidth]{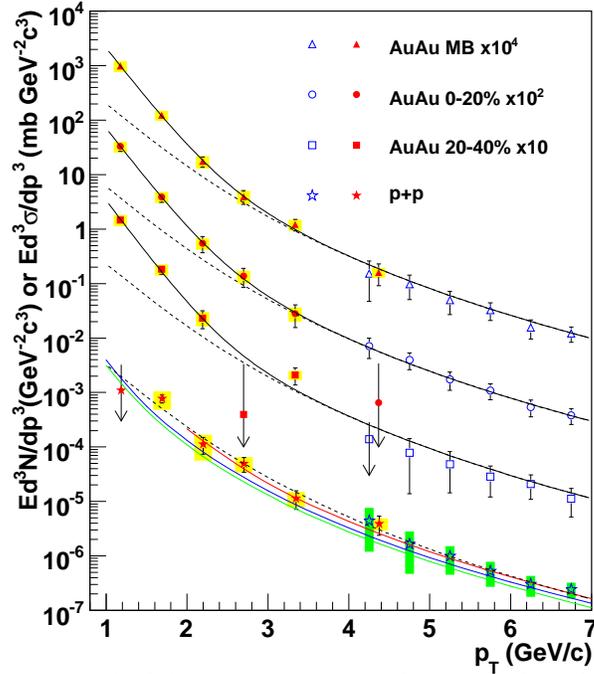}
\end{center}
\vspace*{-1cm}
\caption[]{Invariant yield of direct photons as a function of $p_T$
for $p+p$ and three different centrality classes in Au+Au (solid
symbols). The result of an earlier EMCal measurement is also shown
(open symbols) \cite{Adler:2005ig,Adler:2006yt}.  The $p+p$ data is compared to a pQCD
calculation shown as three lines for different scales.  The dashed
line is a fit of a modified power law to the $p+p$ data.  The Au+Au data
are compared to the $T_{AA}$-scaled $p+p$ fit (dashed line).  The solid
line shows the result of a fit where an exponential is added to the
$T_{AA}$-scaled $p+p$ fit \cite{Adare:2008fqa}.}
\label{fig:yield}
\end{figure}
For $1<p_T<3$ GeV, the direct photon invariant yield in Au+Au is
enhanced above the $T_{AA}$-scaled modified power law that was fit to
the $p+p$ data (Fig. \ref{fig:yield}).  It can be well described,
however, if an exponential is added.  The resulting fit yields
negative inverse exponential slopes of about 220 MeV\footnote{If the $p+p$ data are fit with a pure
power law, $T$ increases by 24 MeV in central events.}.  If the medium
were static, $T$ could be interpreted as its temperature.  For a more
realistic temperature estimate, the data is compared to hydrodynamical
models.  Models that fit the data assume initial temperatures, $T_i$,
between 300 and 600 MeV at formation times, $\tau_0$, between 0.6 and
0.15 fm/c, where the temperature and the formation time are
anti-correlated.  The $221 \pm 23 \mbox{(stat.)} \pm 18 \mbox{(sys.)} \mbox{MeV}$
obtained from the data alone serves as a lower limit.  Even the lower
limit, though, lies above the critical temperature predicted by
lattice calculations.

\section{Conclusion}\label{summary}  
An excess of low-mass $e^{+}$--$e^{-}$ pairs above the hadronic
background was observed at intermediate $p_T$ (1 GeV $< p_T <$ 5 GeV)
for both $p+p$ and Au+Au collisions. The excess can be understood as the
internal conversion of direct photons and shows the expected $1/m$
dependence.  In $p+p$, the direct photon yield is consistent with the
result of a pQCD calculation.  In Au+Au, the yield is much larger.  It
is enhanced above the binary-scaled $p+p$ yield, represented by a
modified power law fit to the $p+p$ data and scaled by $T_{AA}$.  The
Au+Au yield can be well described if an exponential is added to the
binary-scaled $p+p$ fit.  The negative inverse slope of the exponential
is 221 $\pm$ 23 (stat.) $\pm$ 18 (sys.) MeV in central Au+Au.  For a
static medium, this could be interpreted as the temperature.  For an
expanding medium, it serves as a lower limit of the temperature.  It
is well above the critical temperature of about 170 MeV.
Hydrodynamical models that fit the data assume initial temperatures,
$T_i$, between 300 and 600 MeV at formation times, $\tau_0$, between
0.15 and 0.6 fm/c.  Together with the earlier WA98 measurement
\cite{Aggarwal:2000th}, this result can be interpreted as the first experimental
evidence that strongly interacting matter can exceed the Hagedorn
temperature of 170 MeV \cite{Frautschi:1971ij}.

\section*{Acknowledgments}
I thank the RIKEN-BNL Research Center for supporting my work and the
DOE for operating RHIC and PHENIX.

\bibliography{bigsky2009StefanBathe}
\bibliographystyle{bigsky2009}
 
\vfill\eject
\end{document}